\documentclass[11pt,a4paper]{article}

\usepackage{epsfig}
\usepackage{bbm}
\usepackage{amsmath}
\usepackage{amsgen}
\usepackage{amsfonts}
\usepackage{amssymb}
\usepackage{amsbsy}

\def\R{\mathbbm{R}}

\def\ex{{\mathbf{e}_x}}
\def\ey{{\mathbf{e}_y}}
\def\ez{{\mathbf{e}_z}}
\renewcommand{\title}[1]{\topsep=0pt\begin{flushleft}\Large\bf#1\end{flushleft}\vspace{12pt}} 
\renewcommand{\author}[1]{\topsep=0pt\begin{flushleft}\large\rm#1\end{flushleft}} 
\newcommand{\address}[1]{\topsep=0pt\begin{flushleft}\footnotesize\it#1\end{flushleft}\vspace{12pt}} 
\renewcommand{\date}[1]{\topsep=0pt\begin{center}\small#1\end{center}\vspace{12pt}} 

\begin{document}
\title{Characterization and parameterization  of the singular manifold of a
  simple 6-6  Stewart platform} 
\address{{\rm Tiago Charters} \\  Department of Mechanical Engineering\\
Instituto Superior de Engenharia de Lisboa (IPL)\\
Rua Conselheiro Emidio Navarro, 1\\ 1959-007 Lisboa, Portugal \\   {\em and}
Centro de F\'{\i}sica Te\'{o}rica e Computacional, University of Lisbon\\
Complexo Interdisciplinar,
Av.~Prof.~Gama Pinto~2\\
P-1649-003 Lisboa,
Portugal\\
 e--mail: tca@cii.fc.ul.pt \\[1ex]
 {\rm Pedro Freitas}\\ 
Department of Mathematics, Faculdade de Motricidade Humana (TU Lisbon)\\  {\em and}
Group of Mathematical Physics of the University of Lisbon\\ 
Complexo Interdisciplinar,
Av.~Prof.~Gama Pinto~2\\
P-1649-003 Lisboa,
Portugal\\  e--mail: freitas@cii.fc.ul.pt}

\begin{abstract}
This paper presents a  study of  the characterization of the singular
manifold of the six-degree-of-freedom parallel manipulator commonly known as the Stewart
platform. We consider a platform with base vertices in a circle and  for which
the bottom  and top plates
are related by a rotation and a contraction. 
It is shown that in this case the platform is always in a singular configuration
and that the singular manifold can be
parameterized by a scalar parameter.
\end{abstract}

\section{Introduction}

\begin{figure}[ht]
\begin{center} \epsfig{figure=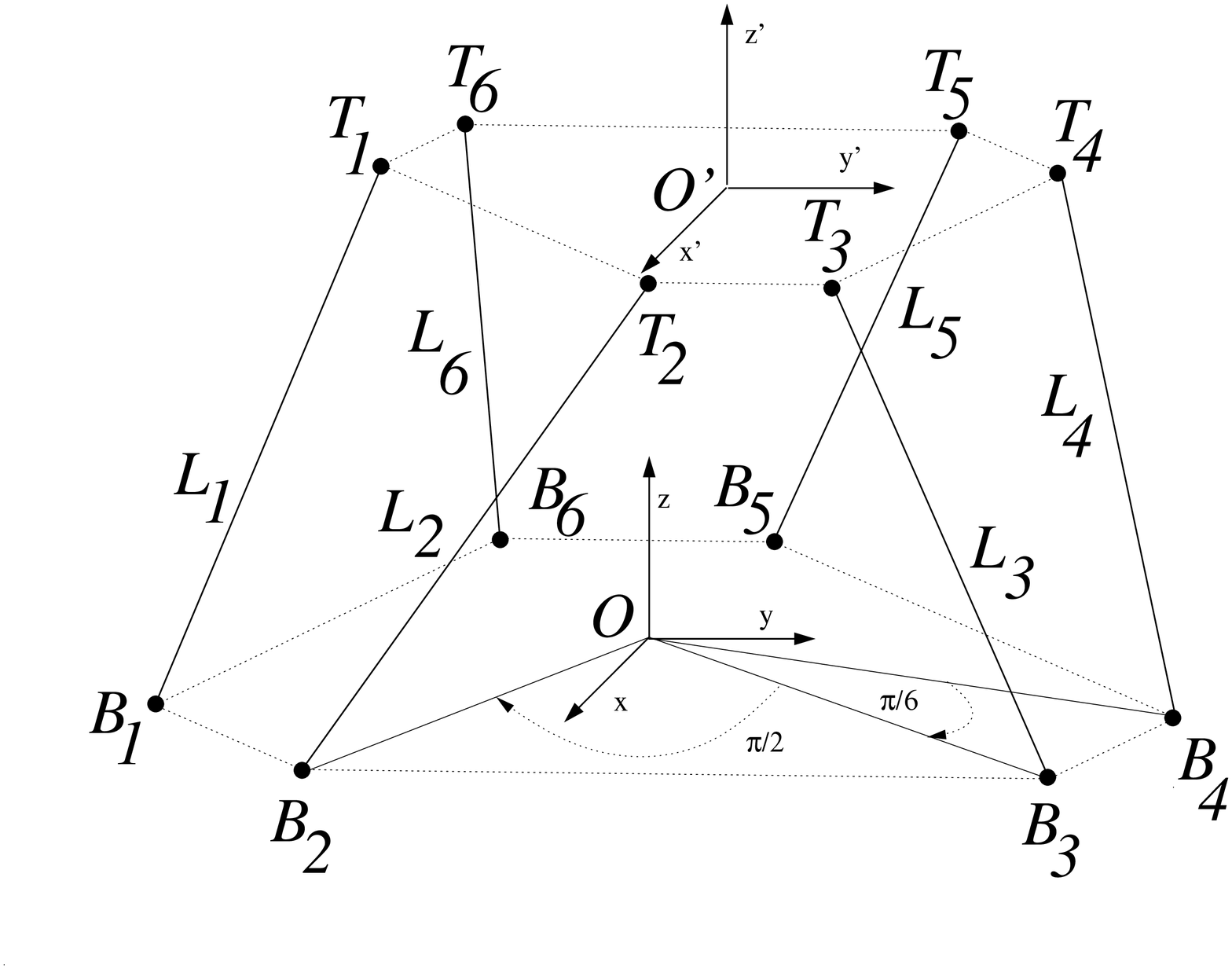,width=8cm}
  \caption{The Stewart platform.}
\label{fig1}
\end{center}
\end{figure}

The Stewart platform is a parallel manipulator with six degrees of freedom \cite{Merlet}.
We will use the (standard) variables  $x, y, z, pitch, roll$ and $yaw$,
where $x,y$ and $z$ are
the coordinates of the centre of the top platform, and $pitch, roll$ and $yaw$ denote the Euler angles
defining the inclination of this platform with respect to the bottom platform,
see Figure \ref{fig1}.

The aim of this paper is to  study the singular manifold which is defined by the
physical  configurations for  which it  will not  be possible  to  determine the
position of the platform  uniquely by fixing the lengths of the  legs. This is a
well-known problem in parallel manipulators\cite{Merlet}.

The solution to the forward kinematics problem naturally divides into two cases,
namely, a  singular and a non-singular.  In the non-singular case  we recall the
work\cite{JiWu01} of  Ji and Wu  and show that  there are $8$  possible isolated
singular solutions  that correspond to the  same legs lengths.   In the singular
case we extend the  previous analysis and show how to obtain,  for a given set of
length legs,  a set of  singular solutions all  of them parameterized by  a scalar
parameter.  These  solutions are  a continuous curves  in position space  and in
rotation space  in which the platform  moves without changing the  values of the
leg lengths. This fully characterize the singular manifold and shows that the
platform is, in this case, completely singular.

Spatial rotations in three dimensions can be parameterized\cite{JiWu01,Diebel06} using both Euler
angles $(\phi, \theta,\psi)$ 
 and unit quaternions $\mathbf{q}=(q_0,q_1,q_2,q_3)$, $\vert\vert
 \mathbf{q}\vert\vert=1$.  A unit quaternion may be described as a vector in
 $\R^4$
\begin{eqnarray}
  \label{eq:quaternion}
  \mathbf{q}&=&(q_0,q_1,q_2,q_3),\\
\mathbf{q}^T\mathbf{q}&=& q_0^2+q_1^2+q_2^2+q_3^2=1.
\end{eqnarray}
The rotation matrix is given by
\begin{eqnarray}
  R=\left(
\begin{matrix}2q_0^2-1+2q_1^2&2q_1q_2-2q_0q_3&2q_0q_2+2q_1q_3\cr 2q_1q_2+2q_0q_3&2q_0^2-1+2q_2^2&2q_2q_3-2q_0q_1\cr
    2q_1q_3-2q_0q_2&2q_0q_1+2q_2q_3&2q_0^2-1+2q_3^2
\end{matrix}
\right).
\end{eqnarray}

Consider the Stewart platform shown
in Figure \ref{fig1}. As shown there, the two coordinate systems $O$ and $O'$ are
fixed to the base and the mobile platforms. The
platform geometry can be described by vectors $\mathbf{L}_i$, $i=1,2,\ldots, 6$,
defined by $\mathbf{L}_i=\mathbf{P}+\mathbf{T}_i-\mathbf{B}_i$,
$i=1,2,\ldots, 6$, where $\mathbf{B}_i$ and $\mathbf{T}_i$ are the base and top
vertices coordinates, respectively, and $\mathbf{P}$ is the center point of the
top plate. We assume that these
points are  related by
\begin{eqnarray}
  \mathbf{T}_i= \mu A \mathbf{B}_i,\qquad i=1,2,\ldots, 6,
\end{eqnarray}
where $A$ is a $3\times 3$ orthogonal matrix ($A^T A=I$, where $I$ is the
 $3\times 3$ identity matrix) and $\mu \in ]0,1[$ is called the rescaling
factor. The coordinates of the base vertices are given by
\begin{eqnarray}
  \mathbf{B}_i=(x_i,y_i,0),\qquad i=1,2,\ldots, 6.
\end{eqnarray}
Given the position $\mathbf{P}=(x,y,z)$ and the transformation matrix $R$
between the two coordinate systems, the leg vectors may be written as
\begin{eqnarray}
  \mathbf{L}_i&=&\mathbf{T}_i-\mathbf{B}_i+\mathbf{P},\\
              &=& (\mu R A -I)\mathbf{B}_i + \mathbf{P},\qquad i=1,2,\ldots, 6.
              \label{eq:Lv}
\end{eqnarray}
So the length for each $i$-leg is given by
\begin{equation}
  \label{eq:L}
  \mathbf{L}_i^{~T}\mathbf{L}_i = \left((\mu R A -I)\mathbf{B}_i + \mathbf{P}\right)^T \left((\mu R A -I)\mathbf{B}_i + \mathbf{P}\right)
\end{equation}

Given $\mathbf{q}$, $A$ and $\mathbf{P}$  the leg lengths are given by
\begin{equation}
\label{eq:Lnorm}
  L_i = \sqrt{\left((\mu R A -I)\mathbf{B}_i + \mathbf{P}\right)^T \left((\mu R A -I)\mathbf{B}_i + \mathbf{P}\right)}.
\end{equation}

\section{Forward kinematics}
In the forward kinematics the six leg lengths $L_i$, $i=1,2,\ldots, 6$, are given, while $R$ and $\mathbf{P}$ are
unknown. Let $\ex=(1,0,0)$, $\ey=(0,1,0)$, $\ez=(0,0,1)$ and expand (\ref{eq:L}), then one gets,
\begin{eqnarray}
  L_i^2&=&\mathbf{P}^T\mathbf{P}+\mathbf{B}_i^{~T}\left((\mu (R A)^T-I)(\mu R A
    -I)\right)\mathbf{B}_i\nonumber\\\qquad 
  && + 2\mathbf{B}_i^{~T}(\mu(R A)^T-I)\mathbf{P},
\end{eqnarray}
or
\begin{eqnarray}
  L_i^2&=&\mathbf{P}^T\mathbf{P}+2x_i\left(\ex^T(\mu (R A)^T \mathbf{P}-\mathbf{P})\right)+2y_i\left(\ey^T(\mu (R A)^T
    \mathbf{P}-\mathbf{P})\right)\nonumber\\\qquad
  && -2\mu\left[x_i^2(\ex^T R A \ex)+x_i y_i(\ex^T R A \ey +\ey^T R A \ex)\right. \nonumber\\\qquad
  &&\left.+y_i^2(\mu\ey^T R A\ey)\right]+(1+\mu^2)(x_i^2+y_i^2).
\label{eq:L_F_K}
\end{eqnarray}
Define $\mathbf{w}=(w_1,w_2,w_3,w_4,w_5,w_6)$ as
\begin{eqnarray}
  w_1&=&\mathbf{P}^T \mathbf{P}\label{eq:w1},\\
  w_2&=& 2\mu\ex^T((R A)^T\mathbf{P}-\mathbf{P})\label{eq:w2},\\
  w_3&=& 2\mu\ey^T((R A)^T\mathbf{P}-\mathbf{P}\label{eq:w3}),\\
  w_4&=& -2\mu\ex^T RA\ex\label{eq:w4},\\
  w_5&=& -2\mu(\left(\ex^T RA\ey +\ey^T RA\ex\right)\label{eq:w5},\\
  w_6&=& -2\mu\ey^T RA\ey\label{eq:w6},
\end{eqnarray}
and $\mathbf{d}=(d_1,d_2,d_3,d_4,d_5,d_6)$, where
\begin{eqnarray}
  \label{eq:d}
  d_i=L_i^2-(1+\mu^2)(x_i^2+y_i^2),\qquad i=1,2,\ldots, 6.
\end{eqnarray}

Then relation (\ref{eq:L_F_K}) can be written as a linear system with the form
\begin{eqnarray}
  \label{eq:ls}
  Q \mathbf{w}=\mathbf{d},
\end{eqnarray}
where the matrix $Q$  is given by
\begin{eqnarray}
\label{eq:Maux}
  Q=\left(\begin{matrix}
1&x_1&y_1&x_1^2&x_1 y_1&y_1^2\cr
1&x_2&y_2&x_2^2&x_2 y_2&y_2^2\cr
1&x_3&y_3&x_3^2&x_3 y_3&y_3^2\cr
1&x_4&y_4&x_4^2&x_4 y_4&y_4^2\cr
1&x_5&y_5&x_5^2&x_5 y_5&y_5^2\cr
1&x_6&y_6&x_6^2&x_6 y_6&y_6^2\cr
\end{matrix}
\right).
\end{eqnarray}
Note  that if the base points are all different and belong to a conic section  then $\det Q =0$.
The  matrix given by (\ref{eq:Maux}) corresponds to the well known Braikenridge-Maclaurin  construction.

In the next sections we will show that one can obtain the rotation matrix  $R$
and the position $\mathbf{P}$ in terms of 
the solution $\mathbf{w}=(w_1,w_2,\ldots,w_6)$ of the linear system given by
(\ref{eq:ls}). The solution to the forward 
kinematics problem naturally divides into two cases, namely, a non-singular case
where $\det Q\ne 0$ and a  singular case where $\det Q=0$.

In the singular case, we
obtain for a given set of length legs, $L_1,L_2,\ldots,L_6$, a singular  solution parameterized by a scalar
parameter. These solutions are curves in position space and in rotation space in
which the platform moves without changing the values of the leg lengths.

\subsection{Non-singular case}
In the case where the six base vertices are not on a conic section, one gets
$\det Q\ne 0$,   and so the solution of (\ref{eq:ls}),
$\mathbf{w}=(w_1,w_2,w_3,w_4,w_5)$, can be obtained from 
\begin{eqnarray}
  \mathbf{w}=Q^{-1}d.
\end{eqnarray}
The first three equations (\ref{eq:w1}), (\ref{eq:w2}) and (\ref{eq:w3})
determines the rotation parameters, namely, 
$\mathbf{q}$, and the last three (\ref{eq:w4}), (\ref{eq:w5}) and (\ref{eq:w6})
the position $\mathbf{P}=(x,y,z)$.

To determine the rotation parameters consider the equations
\begin{eqnarray}
  w_4&=&-2 \mu \left(2 { q_1}^2+2 { q_0}^2-1\right),\\
  w_5&=&-8 \mu { q_1} { q_2}\label{eq:q1q2}, \\
  w_6&=&-2 \mu \left(2 { q_2}^2+2 { q_0}^2-1\right),
\end{eqnarray}
which are obtained from (\ref{eq:w4}), (\ref{eq:w5}) and (\ref{eq:w6}), respectively.
Eliminating $q_0$, one gets,
\begin{eqnarray}
 { q_1}^2- {q_2}^2&=&-(w_4-w_6)/(4\mu),\\
 { q_1} { q_2} &=&-w_5/(8\mu).
\end{eqnarray}
Let 
\begin{eqnarray}
  \alpha=\frac{w_4-w_6}{4\mu},\qquad  \beta= -\frac{w_5}{8\mu}.
\end{eqnarray}
Then the above equations can be written as
\begin{eqnarray}
  q_1^4+\alpha q_1^2-\beta^2&=&0,\\
  q_2^4-\alpha q_2^2-\beta^2&=&0.
\end{eqnarray}
So,
\begin{eqnarray}
  q_1^2&=& \frac{-\alpha+\gamma}{2},\label{eq:q1sol}\\
  q_2^2&=& \frac{\alpha+\gamma}{2},\label{eq:q2sol}
\end{eqnarray}
where
\begin{eqnarray}
  \gamma=\sqrt{\alpha^2+4\beta^2}.
\end{eqnarray}
Substituting yields
\begin{eqnarray}
  q_3^2&=& \frac{1}{2} + \frac{w_4}{4\mu}-\frac{\alpha+\gamma}{2},\label{eq:q3sol}\\
  q_0^2&=&\frac{1}{2}- \frac{w_4}{4\mu}+\frac{\alpha-\gamma}{2}.\label{eq:q0sol}
\end{eqnarray}
Assuming $q_0\ge 0$ and that (\ref{eq:q3sol}) and (\ref{eq:q2sol}) have  two
roots each, then, $q_1$ is determined by (\ref{eq:q1q2}).
Consequently, we  have a total of four different quaternions. These are
\begin{eqnarray}
  \mathbf{s_1}&=&(\bar q_0,\bar q_1,\bar q_2,\bar q_3),\\
  \mathbf{s_2}&=&(\bar q_0,\bar q_1,\bar q_2,-\bar q_3),\\
  \mathbf{s_3}&=&(\bar q_0,-\bar q_1,-\bar q_2,\bar q_3),\\
  \mathbf{s_4}&=&(\bar q_0,-\bar q_1,-\bar q_2,-\bar q_3),
\end{eqnarray}
where $(\bar q_0,\bar q_1,\bar q_2,\bar q_3)$ are the roots.


To determine the position, consider the equations
\begin{eqnarray}
  \mathbf{u}^T&=&2\mu\ex^T((R A)^T-I),\\
  \mathbf{v}^T&=&2\mu\ey^T((R A)^T-I).
\end{eqnarray}
Thus
\begin{eqnarray}
  \mathbf{P}^T\mathbf{P}&=&w_1,\label{eq:sphere}\\
  \mathbf{u}^T\mathbf{P} &=&w_2,\label{eq:plane1}\\
  \mathbf{v}^T\mathbf{P} &=&w_3,\label{eq:plane2}
\end{eqnarray}
Obviously (\ref{eq:plane1}) and (\ref{eq:plane2}) represent two planes and their
intersection is a line with equation given by
\begin{eqnarray}
\label{eq:line}
  \mathbf{P}=\mathbf{r}_0+t \mathbf{r}_1,
\end{eqnarray}
where $t$ is the parameter of the line. The vectors $\mathbf{r}_0$ and $\mathbf{r}_1$ are given by
\begin{eqnarray}
  \mathbf{r}_0&=&\frac{(\mathbf{v}^T\mathbf{v})w_2-(\mathbf{u}^T\mathbf{v})w_3}{(\mathbf{u}^T
    \mathbf{u})(\mathbf{v}^T\mathbf{v})-(\mathbf{u}^T\mathbf{v})^2}\mathbf{u}-
\frac{-(\mathbf{u}^T\mathbf{v})w_2+(\mathbf{u}^T\mathbf{u})w_3}{(\mathbf{u}^T
    \mathbf{u})(\mathbf{v}^T\mathbf{v})-(\mathbf{u}^T\mathbf{v})^2}\mathbf{v},\\
  \mathbf{r}_1&=&\frac{\mathbf{u}\times \mathbf{v}}{\vert\vert\mathbf{u}\times \mathbf{v}\vert\vert}.
\end{eqnarray}
The line (\ref{eq:line}) intersects the sphere (\ref{eq:sphere}) at two points $P_\pm$  given by
\begin{eqnarray}
  P_\pm&=& \mathbf{r}_0\pm t^*\mathbf{r}_1,\\
\end{eqnarray}
where
\begin{eqnarray}
  t^*=\sqrt{w_1-\mathbf{r}_0^T\mathbf{r}_0}.
\end{eqnarray}
Note that in order to $P_\pm$ exist one should have
\begin{eqnarray}
  \label{eq:t*}
  w_1\ge \mathbf{r}_0^T\mathbf{r}_0.
\end{eqnarray}
So, both $R$ and $\mathbf{P}$ are found, and totally they have eight possible
different solutions for a given set of leg lengths.

\subsection{Singular case\label{sing}}
In this case, we assume that all points belong to a circle $x_i^2+y_i^2=1$ (we can assume $r=1$ without loss of
generality), $i=1,2,\ldots,6$. In this case the matrix
\begin{eqnarray}
  \label{eq:QCirc}
  Q=\left(\begin{matrix}
1&x_1&y_1&x_1^2&x_1 y_1&1-x_1^2\cr
1&x_2&y_2&x_2^2&x_2 y_2&1-x_2^2\cr
1&x_3&y_3&x_3^2&x_3 y_3&1-x_3^2\cr
1&x_4&y_4&x_4^2&x_4 y_4&1-x_4^2\cr
1&x_5&y_5&x_5^2&x_5 y_5&1-x_5^2\cr
1&x_6&y_6&x_6^2&x_6 y_6&1-x_6^2\cr
\end{matrix}\right)
\end{eqnarray}
is singular, that is, $\det Q=0$ and in fact, if all points are different and
belong to a conic
section the rank of $Q$ is five (corresponding to the Braikenridge-Maclaurin
construction). This will be the case if 
$x_i^2+y_i^2=1$, $i=1,2,\ldots,6$, and $(x_i,y_i)\ne(x_j,y_j)$ for $i\ne j$, $i,j=1,2,\ldots,6$.

This fact enables us to explicitly  compute the $LU$ factorization of the matrix
$Q$  in  terms of  the  coordinate  of the  vertices  of  the base  $(x_i,y_i)$,
$i=1,2,\ldots,6$. These expressions are to big to be shown here but a script for
the Maxima computer algebra system\cite{Maxima:CAS} is available upon request to the author.

So the linear system $Q\mathbf{w}=d$ can be put into the for
\begin{eqnarray}
  \label{eq:lnLU}
  U\mathbf{w}=L^{-1}d,
\end{eqnarray}
where $\det L=1$ and $U$ is a matrix with rank $5$. The solution of
(\ref{eq:lnLU}) is given in terms of a 
 solution $(w_2,w_3,w_4,w_5,w_6)$ which depends on the value of $w_1$, which 
 we take to be a free parameter. Notice that any other quantity could be used
 for this purpose, although expression (\ref{eq:sphere}) suggests that $w_1$ is
 the good choice. So the 
 expressions given by (\ref{eq:q1sol}), (\ref{eq:q2sol}), (\ref{eq:q3sol}) and
 (\ref{eq:q0sol}) can be used to determine
 the the values of the quaternion $\mathbf{q}$, the rotation matrix, and the
 point $\mathbf{P}$ as a function of the
 free parameter $w_1$. 

\section{Conclusions} 
The singular  manifold of a Stewart platform  is define by
the physical configurations  for which it will not be  possible to determine the
position  of the  platform  uniquely by  fixing  the lengths  of  the legs.   By
considering a  simple Stewart  platform, for  which the base  vertices are  in a
circle (although the result  naturally holds for any conic section)
  and  the  bottom  and  top  plates  are related  by  a  rotation  and  a
contraction, it was shown that the platform is always in a singular configuration.
It was also   shown  how to  characterize the singular manifold in this case  and how it
can be parameterize by a scalar parameter.


\begin{thebibliography}{9999}

\bibitem{Merlet}   Merlet, J. P., {\em Parallel Robots}, Springer, (2006)

\bibitem{JiWu01} Ji, P., and Wu H., \emph{A Closed-Form Kinematics Solutions for
  the $6-6^p$ Stewart Platform},
IEEE Transactions on robotics and automation, Vol. 17, 4, (2001), 522-526
 
\bibitem{Diebel06} Diebel, J., \emph{Representing Attitude: Euler Angles, Quaternions, and Rotation Vectors}\\
{\small http://ai.stanford.edu/{\~\,}diebel/attitude/attitude.pdf}

\bibitem{Maxima:CAS} http://maxima.sourceforge.net
\end{thebibliography}
\end{document}